 \documentclass[12pt,letterpaper]{article}
 \usepackage{jheppub}
 \usepackage{cancel}
\usepackage{hyperref}

\usepackage{esint}
 
 \def\bR{{\mathbb R}}

 \newcommand{\vev}[1]{{\left< {#1} \right>}} 
\newcommand{\be}{\begin{equation}}
\newcommand{\ee}{\end{equation}}
\newcommand{\galg}{\mathfrak{g}}

\title{Probing ${\cal N}=2$ superconformal field theories with localization}

\author[a]{Bartomeu Fiol,}
\author[b]{Blai Garolera }
\author[a]{ and Gen\'is Torrents}

\affiliation[a]{Departament de F{\'\i}sica Fonamental i \\Institut de Ci{\`e}ncies del Cosmos, 
Universitat de Barcelona,
Mart{\'\i}\ i Franqu{\`e}s 1, 08028 Barcelona, Catalonia, Spain }

\affiliation[b]{Escuela de F{\'\i}sica,
Universidad de Costa Rica, 11501-2060 San Jos\'e,
Costa Rica}

\emailAdd{bfiol@ub.edu}
\emailAdd{blai.garolera@ucr.ac.cr}
\emailAdd{genistv@icc.ub.edu }

\abstract{We use supersymmetric localization to study probes of four dimensional Lagrangian ${\cal N}=2$ superconformal field theories. We first derive a unique equation for the eigenvalue density of these theories. We observe that these theories have a Wigner eigenvalue density precisely when they satisfy a necessary condition for having a holographic dual with a sensible higher-derivative expansion. We then compute in the saddle-point approximation the vacuum expectation value of 1/2-BPS circular Wilson loops, and the two-point functions of these Wilson loops with the Lagrangian density and with the stress-energy tensor. This last computation also provides the corresponding Bremsstrahlung functions and entanglement entropies. As expected, whenever a finite fraction of the matter is in the fundamental representation, the results are drastically different from those of ${\cal N}=4$ supersymmetric Yang-Mills theory.
}  

\begin{document}
\maketitle
\section{Introduction}
Studying the response of quantum field theories to the presence of external probes is an interesting way to understand these theories better. However, for generic quantum field theories it is prohibitively hard to obtain exact results. The situation improves for theories with additional symmetries, like conformal invariance and/or supersymmetry. In particular, in conformal field theories, for simple enough questions, the additional symmetry disentangles the space-time dependence from the coupling dependence, and the full answer is given in terms of some unknown coefficients that possibly depend on the marginal couplings of the CFT \cite{Kapustin:2005py}. In these cases, to actually compute these coefficients, one must resort to other techniques to determine them, like perturbation theory, the AdS/CFT correspondence, integrability or supersymmetric localization.

Besides their intrinsic interest, a comparatively less explored but potentially far-reaching application of the study of probes in conformal field theories is as useful diagnostics to characterize their holographic duals \cite{Rey:2010ry, Passerini:2011fe, Fraser:2011qa}. 

In this note we are going to focus on the study of probes in the fundamental representation of four dimensional ${\cal N}=2$ conformal gauge theories. The main reason to limit ourselves to this small family of conformal field theories is to take full advantage of the technique of supersymmetric localization \cite{Pestun:2007rz}. There have been already many works devoted to the use of supersymmetric localization to study probes of these theories (see \cite{Okuda:2014fja} for a review). The main novelties of the present work are the derivation of a single integral equation that governs the eigenvalue density of all these SCFT, in the saddle-point approximation, and the study of correlators of Wilson loops with local operators. Let's comment on these two points in some detail. 

The matrix models that compute the partition functions of all these superconformal theories can't be solved exactly at finite N (except for the case of ${\cal N}=4$ theories). We resort to study their partition functions in the saddle-point approximation, by introducing eigenvalue densities $\rho(x)$ for each of them. As a first result, we notice that we can write the integral equation for the eigenvalue densities of all these theories in a unified way
\be
\int_{-\mu}^\mu dy \rho(y)\left(\frac{1}{x-y}-\nu K(x-y)\right)=\frac{8\pi^2}{ \lambda}x-\nu K(x) \, ,
\label{integralintro}
\ee
where $K(x)$ is a function to be defined below and the parameter $\nu$ counts what fraction of the matter multiplets transforms in the fundamental representation of the gauge group. Roughly speaking, the large N limit washes out many finite N details about the gauge groups and the relevant representations, and the only possible contributions of hypermultiplets in the fundamental representation to the matter content of the theory are $\nu=0,\frac{1}{2},1$.

Even before we attempt to solve equation (\ref{integralintro}), it is apparent that the resulting eigenvalue density presents two qualitatively different behaviors, for $\nu=0$ and $\nu>0$. This was already realized in \cite{Rey:2010ry}, where the cases $\nu=0,1$ were compared.
In \cite{Rey:2010ry} it was argued as well that the physical reason for the qualitatively different behavior are the screening properties of matter in the fundamental representation.

It is also worth pointing out that this sharply different behavior has a reflection on the possible holographic duals of these field theories. It was argued in \cite{Buchel:2008vz} that a necessary condition for a 4d CFT to have a holographic dual described by a gravitational action with a sensible higher derivative expansion is that at large N their central charges satisfy
\begin{equation}
c,a\gg 1,  \hspace{1cm}
\frac{|c-a|}{c}\ll 1 \,.
\label{bucheletal}
\end{equation}
As it turns out, among the ${\cal N}=2$ SCFTs considered here, only theories with $\nu=0$ (i.e. the number of hypermultiplets in the fundamental representation does not scale with $N$) satisfy this constraint. So we observe a correlation between having a Wigner eigenvalue density and potentially having a holographic dual with a sensible derivative expansion.

Turning to the solution of eq. (\ref{integralintro}), in the limit of strictly infinite 't Hooft coupling, $\lambda=g^2_{YM}N$, we find an analytic expression for the eigenvalue density, slightly generalizing the result in \cite{Passerini:2011fe}. For strong but finite 't Hooft coupling, we can't solve analytically the saddle point equation, so we must resort to some approximation. We do so by following a couple of methods already present in the literature \cite{Passerini:2011fe, Bourgine:2011ie}.

Once we have found the eigenvalue density for generic $\nu$, we put this result to use by computing various correlation functions involving circular Wilson loops in the fundamental representation. The qualitatively different behavior mentioned above is manifested here as follows: for $\nu=0$ theories like ${\cal N}=4$ SYM, the vev grows exponentially in $\sqrt{\lambda}$ \cite{Berenstein:1998ij}, while for $\nu =1$ theories, like ${\cal N}=2$ SQCD, it grows with a power law, $\vev{W}\sim \lambda^3$ \cite{Passerini:2011fe} . One can then anticipate that for theories with $\nu=1/2$, the vev of the circular Wilson loop should present a growth in between $\nu=0$ and $\nu=1$. Indeed, we obtain
\be
\vev{W}_{\nu=\frac{1}{2}}\sim \lambda^5 \,.
\ee

We then compute the two-point function of this Wilson loop with local operators. More specifically, we compute the normalized two-point function of the straight Wilson line with the Lagrangian density, and then with the stress-energy tensor. Conformal invariance fixes these normalized two-point functions up to a single coefficient each. In ${\cal N}=4$ SYM, the coefficients are essentially the same \cite{Fiol:2012sg}, since the Lagrangian density and the stress-energy tensor belong to the same supermultiplet. This is no longer the case for ${\cal N}=2$ SCFTs, so it is interesting to obtain and compare the behavior of these coefficients. We find that for these two-point functions the $\nu\neq 0$ dependence enters through an angle $\theta$ defined by
\be
\cos \theta=1-\nu \, .
\ee
For instance, the coefficient in the two-point function of the straight Wilson line with the Lagrangian density is given by a constant in the large N, large $\lambda$ limit
\be
\frac{\vev{{\cal L}(x) W}}{\vev{W}}= \frac{f_W}{|\vec x|^4}\ , \hspace{1cm} f_W=\frac{1}{8\pi^2}\left(\frac{2\pi}{\theta}-1 \right) \, .
\ee
Recently, it has been conjectured in \cite{Fiol:2015spa} that for ${\cal N}=2$ SCFTs, one can compute the normalized two-point function of the straight Wilson line with the stress-energy tensor from the vev of the Wilson loop in a squashed four-sphere, $\mathbb{S}^4_b$. Granting that this conjecture is correct, this two-point function displays a logarithmic dependence on the coupling 
\be 
\frac{\vev{T_{00}(x) W}}{\vev{W}}=\frac{h_W}{|\vec x|^4} \,  , \hspace{1cm} h_W=\frac{1}{6\pi \theta} \ln \lambda \, .
\ee
Furthermore, based on \cite{Lewkowycz:2013laa} it was also conjectured in \cite{Fiol:2015spa} that the Bremsstrahlung function \cite{Correa:2012at} of this Wilson loop for any ${\cal N}=2$ SCFT is given by essentially the same coefficient above, 
\be
B= 3h_W=\frac{1}{2\pi \theta} \ln \lambda \,.
\ee
Finally, using the general expression derived in \cite{Lewkowycz:2013laa}, we also compute the change in entanglement entropy of a spherical region of the vacuum state, due the presence of these probes. It is given by
\be
S=\left(\frac{2\pi}{3\theta}-1\right)  \ln \lambda \,.
\ee

These results are to be contrasted with the well-known corresponding results for ${\cal N}=4$ SYM. In this case, all these coefficients are essentially the same due to the extra amount of supersymmetry, $B=3 h_W=4 f_W$, and can be computed exactly \cite{Correa:2012at, Fiol:2012sg} for various gauge groups and representations \cite{Fiol:2013hna, Fiol:2014fla}. In the large N, large $\lambda$ regime, they scale as $\sqrt{\lambda}$. Our results further exemplify to what extent the properties of probes of ${\cal N}=4$ SYM are not generic among ${\cal N}=2$ theories.

As possible extensions of this work, our results could be generalized to Wilson loops in higher rank representations. It might be also interesting to compute subleading corrections to the results obtained here. Finally, while we carry out this analysis for Lagrangian theories for which $\nu$ can only take the values $\nu=0,\frac{1}{2},1$, an interesting question is whether there are non-Lagrangian ${\cal N}=2$ SCFTs whose correlators are captured by the expressions presented here, for other values of $\nu$.

The present paper is organized as follows: in section 2 we introduce the superconformal theories that we are going to study, we recall the matrix model that computes their partition function, and derive their eigenvalue density in the large N, large $\lambda$ regime. In section 3 we use this eigenvalue density to compute various correlation functions related to heavy probes coupled to these theories.

{\bf Note added:} As this paper was being typed, we learned of upcoming work \cite{mitevpomoni} that studies similar matters for quiver ${\cal N}=2$ SCFTs.  In that work, the regime when one of the gauge couplings is strong while the other ones tend to zero is not considered, so there is no immediate overlap with the present paper.

\section{Saddle-point equation for  4d ${\cal N}=2$ SCFTs}
In this section we present the ${\cal N}=2$ superconformal field theories (SCFTs) that we are going to study and recall the matrix model that computes their partition functions. We then derive the saddle-point equation for these matrix models, and solve them in the large N, large $\lambda$ limit, to obtain their eigenvalue densities.

Let's start by recalling how to obtain all 4d ${\cal N}=2$ SCFTs theories, with a single gauge group, and a marginal coupling. With ${\cal N}=2$ supersymmetry the $\beta$-function is exactly zero if and only if the one-loop contribution is zero \cite{Howe:1983wj}. Since we are interested in SCFTs that admit a large N limit, we restrict to classical (i.e. non-exceptional) gauge groups and matter content in representations with up to two indices: fundamental, 2-symmetric, 2-antisymmetric and adjoint. The complete list of such theories is well-known \cite{Koh:1983ir}, and we present it in table \ref{tab:list}, together with their central charges.

A quantity that will turn out to be relevant in what follows is 
\be
\nu \equiv \lim_{N\rightarrow \infty} \frac{n_f}{2N}
\ee
which counts what fraction of the matter in these theories belongs to the fundamental representation in the large N limit. For these theories, we observe in table \ref{tab:list} that  $\nu$ can only take the values $\nu=0,1/2,1$.

\begin{table}
\begin{center}
    \begin{tabular}{| c | c | c | c | c |}
    \hline
    \multicolumn{5}{|c|}{$SU(N)$} \\ \hline
    $(n_{adj},n_f,n_{S_2},n_{A_2})$ & $c$ & $a$ & $\delta\equiv(c-a)/c$ & $\nu$ \\ \hline
    $(1,0,0,0)$ & $\frac{1}{4}N^2-\frac{1}{4}$ & $\frac{1}{4}N^2-\frac{1}{4}$ & $0$ & 0 \\ \hline
    $(0,0,1,1)$ & $\frac{1}{4}N^2-\frac{1}{6}$ & $\frac{1}{4}N^2-\frac{5}{24}$ &  $\frac{1}{6N^2}+\mathcal{O}(N^{-4})$ & 0 \\ \hline
    $(0,4,0,2)$ & $\frac{1}{4}N(N+1)-\frac{1}{6}$ & $\frac{1}{4}N(N+\frac{1}{2})-\frac{5}{24}$ & $\frac{1}{2N}+\mathcal{O}(N^{-2})$ & 0 \\ \hline
    $(0,2N,0,0)$ & $\frac{1}{3}N^2-\frac{1}{6}$& $\frac{7}{24}N^2-\frac{5}{24}$ & $\frac{1}{8}+\mathcal{O}(N^{-2})$ & 1 \\ \hline
    $(0,N+2,0,1)$ & $\frac{7}{24}N^2+\frac{1}{8}N-\frac{1}{6}$ & $\frac{13}{48}N^2+\frac{1}{16}N-\frac{5}{24}$ & $\frac{1}{14}+\mathcal{O}(N^{-1})$ & $\frac{1}{2}$ \\ \hline
    $(0,N-2,1,0)$ & $\frac{7}{24}N^2-\frac{1}{8}N-\frac{1}{6}$ & $\frac{13}{48}N^2-\frac{1}{16}N-\frac{5}{24}$ & $\frac{1}{14}+\mathcal{O}(N^{-1})$ & $\frac{1}{2}$ \\ \hline
    \multicolumn{5}{c}{} \\ \hline
    \multicolumn{5}{|c|}{$SO(2N)$} \\ \hline
    $(n_{adj},n_f,n_{S_2})$ & $c$ & $a$ & $\delta\equiv(c-a)/c$ & $\nu$ \\ \hline
    $(1,0,0)$ & $\frac{1}{2}N^2-\frac{1}{4}N$ & $\frac{1}{2}N^2-\frac{1}{4}N$ & 0 & 0 \\ \hline
    $(0,2N-2,0)$ & $\frac{2}{3}N^2-\frac{1}{2}N$ & $\frac{7}{12}N^2-\frac{3}{8}N$ & $\frac{1}{8}-\frac{3}{32 N}+\mathcal{O}(N^{-2})$ & 1 \\ \hline 
    \multicolumn{5}{c}{} \\ \hline
    \multicolumn{5}{|c|}{$SO(2N+1)$}  \\ \hline
    $(n_{adj},n_f,n_{S_2})$ & $c$ & $a$ & $\delta\equiv(c-a)/c$ & $\nu$ \\ \hline    
    $(1,0,0)$ & $\frac{1}{2}N^2+\frac{1}{4}N$ & $\frac{1}{2}N^2+\frac{1}{4}N$ & 0 & 0 \\ \hline
    $(0,2N-1,0)$ & $\frac{2}{3}N^2+\frac{1}{6}N-\frac{1}{12}$ & $\frac{7}{12}N^2+\frac{5}{24}N-\frac{1}{24}$ & $\frac{1}{8}-\frac{3}{32 N}+\mathcal{O}(N^{-2})$  & 1 \\ \hline 
    \multicolumn{5}{c}{} \\ \hline
    \multicolumn{5}{|c|}{$Sp(2N)$}  \\ \hline
    $(n_{adj},n_f,n_{A_2})$ & $c$ & $a$ & $\delta\equiv(c-a)/c$ & $\nu$ \\ \hline    
    $(1,0,0)$ & $\frac{1}{2}N^2+\frac{1}{4}N$ & $\frac{1}{2}N^2+\frac{1}{4}N$ & 0 & 0 \\ \hline
    $(0,4,1)$ & $\frac{1}{2}N^2+\frac{3}{4}N-\frac{1}{12}$ & $\frac{1}{2}N^2+\frac{1}{2}N-\frac{1}{24}$ & $\frac{1}{2N}+\mathcal{O}(N^{-2})$ & 0 \\ \hline
    $(0,2N+2,0)$ & $\frac{2}{3}N^2+\frac{1}{2}N$ & $\frac{7}{12}N^2+\frac{3}{8}N$ & $\frac{1}{8}+\frac{3}{32 N}+\mathcal{O}(N^{-2})$ & 1 \\
    \hline
    \end{tabular}
\end{center}
\caption{List of 4d ${\cal N}=2$ SCFT families admitting a large N limit for each classical simple Lie algebra } \label{tab:list}
\end{table}

\subsection{Partition function}
Due to supersymmetric localization, the partition function of these theories on $\mathbb{S}^4$ reduces to an integral over the Lie algebra $\mathfrak{g}$ of the gauge group
 \cite{Pestun:2007rz}
\be
Z_{\mathbb{S}^4}=\frac{1}{\mbox{vol}(G)}\int_{\galg}[da]e^{-\frac{8\pi^2r^2}{g^2_{YM}}(a,a)}Z_{1-loop}(ra)|Z_{inst}(ia,r^{-1},r^{-1},q)|^2 \, ,
\ee
where $(\, , \, )$ denotes the bilinear form obtained from tracing the product in the fundamental representation and $r$ is the radius of $\mathbb{S}^4$. 
 This formula can be rewritten in terms of an integral over the Cartan subalgebra whose integration measure is given by a Faddeev-Popov determinant of the form 
$$
 \Delta^{2}(a) = \prod_{\alpha\in roots({\galg})}(\alpha\cdot a)^{2}\, .
$$
In this gauge the factor $Z_{1-loop}(ra)$ is a certain infinite dimensional product, which appears as a 1-loop determinant in the localization computation.
For an $\mathcal{N}=2$ theory with massless hypermultiplets in any $G$-representation $\mathcal{R}$, the 1-loop determinant is \cite{Pestun:2007rz}
\be
Z^{\mathcal{N}=2,W}_{1-loop}(ra)=\frac{\prod_{\alpha\in roots({\galg})}H( \alpha\cdot a r)}{\prod_{w\in weights(\mathcal{R})}H(w \cdot a r)} \, ,
\label{zoneloop}
\ee
where $H(x)$ is given by
$$
H\left(x\right)=\prod_{n=1}^{\infty}\left(\left(1+\frac{x^{2}}{n^{2}}\right)^{n}e^{-x^{2}/n}\right) \, .
$$
Formula (\ref{zoneloop}) literally holds if the divergent factors are the same in the one-loop determinants for the vector and hypermultiplets. This happens for representations $W$ such that
\be
\sum_{\alpha}(\alpha \cdot a)^2=\sum_w(w \cdot a)^2 \ ; \ \ a \in {\galg} \nonumber
\ee
that is if the $\beta$-function vanishes and the $\mathcal{N}=2$ theory is superconformal.

The factor $Z_{inst}(ia,\epsilon_1,\epsilon_2,q)$ is the Nekrasov's instanton partition function of the gauge theory in the $\Omega$-background on $\bR^4$ \cite{Nekrasov:2002qd}. For $\mathcal{N}=4$ all instanton corrections vanish ($Z_{inst}=1$). As is customary, we will assume that their contribution is negligible in the large N limit.


We now proceed to derive the saddle-point equation for these matrix models.
Following the standard procedure, we bring the Faddeev-Popov and one-loop factors to the exponent. In the large N limit, we can pass to a continuum version. To do so, introduce the eigenvalue density
\be
\rho(x)=\frac{1}{N}\sum_i\delta(x-ra_i)\, ,
\ee
defined in the interval $\Gamma =[-\mu,\mu]$ and unit normalized. It is convenient to introduce
\be
K(x)=-\frac{d\, \log H(x)}{d\, x} \, .
\ee
Since $H(x)$ is an even function under $x\rightarrow -x$, $K(x)$ is odd. It is straightforward to write down an integral equation for the eigenvalue density for each ${\cal N}$=2 SCFT. We are now going to argue that all these integral equations can be written in a unified fashion. Let's first consider SCFTs with gauge group $SU(N)$, and for concreteness let's illustrate the argument with the specific example of the SCFT with a hypermultiplet in the antisymmetric representation and $N+2$ hypermultiplets in the fundamental one. The integral equation for the eigenvalue density is
\be
\int _{-\mu}^{\mu}dy\rho(y)\left(\frac{1}{x-y}-K(x-y)+\frac{1}{2}K(x+y)\right)=\frac{8\pi^2}{\lambda}x-\frac{1}{2}K(x) \, .
\ee
The terms inside the parenthesis in the integral come respectively from the Faddeev-Popov determinant, the vector multiplet contribution and the hypermultiplet in the antisymmetric representation. The $K(x)$ term on the RHS corresponds to the hypermultiplets in the fundamental representation. Combining this equation with the one that we obtain by changing $x\rightarrow -x, y\rightarrow -y$, we learn that the eigenvalue density is even. Then, by combining this equation with the one we obtain by changing $x\rightarrow -x$ we learn that under the integral $K(x+y)$ can be replaced by $K(x-y)$. The same argument goes through for all the other SCFTs with gauge group SU(N), and we learn that their integral equations can be written in a compact way in terms of $\nu$
\be
\int_{-\mu}^\mu dy \rho(y)\left(\frac{1}{x-y}-\nu K(x-y)\right)=\frac{8\pi^2}{\lambda}x-\nu K(x) \, .
\ee
The discussion can be easily generalized to SCFTs with other classical gauge groups. For $G=SO(2N)$ the Faddeev-Popov determinant is
\be
\Delta^2(a)=\prod_{i<j}^N|a_i^2-a_j^2|^2 \, ,
\ee
so its contribution to the integral equation is naively different from the case of $SU(N)$. However
\be 
\int_{-\mu}^{\mu}dy \rho(y)\frac{2x}{x^2-y^2}=\int_{-\mu}^{\mu}dy \rho(y) \frac{2}{x-y}-\cancelto{0}{\int_{-\mu}^{\mu}dy \rho(y)\frac{2y}{(x-y)(x+y)}} \, ,
\ee
so it turns out to give the same kernel as $SU(N)$, except for a factor of two. A factor of two will be generated as well in the term with $\lambda^{-1}$ because the trace in the fundamental representation includes both $\pm a_i$ weights for $SO(2N)$. Finally, the Faddeev-Popov determinants of $SO(2N+1),Sp(2N)$ present further additional terms, but their contribution is subleading in the large N limit. The upshot of this analysis is that for all these SCFTs, the singular integral equation that determines the eigenvalue distribution is
\begin{equation}
\int_{-\mu}^\mu dy \rho(y)\left(\frac{1}{x-y}-\nu K(x-y)\right)=\frac{8\pi^2}{
\lambda}x-\nu K(x) \, ,
\label{thesaddleeq}
\end{equation}

For $\nu=0$ this is of course the integral equation for the Wigner distribution, while for $\nu=1$ this equation was derived in \cite{Passerini:2011fe} for the particular case of ${\cal N}=2$ SQCD. 

Before we proceed, let's pause to comment on the holographic implications of this result. A very interesting question is what 4d CFTs admit a holographic dual with a sensible gravitational description in at least some regime of parameters. In this regard, it is possible to find necessary conditions in terms of the central charges of the 4d CFT. If one requires that the gravitational dual is described by an action with two derivatives (i.e. Einstein -Hilbert in the gravitational sector) then in the large N limit the central charges must satisfy \cite{Henningson:1998gx}
\begin{equation}
c,a \gg1 \, , \hspace{1cm} c-a=0+{\cal O}(1/N) \, .
\label{hensken}
\end{equation}
If one relaxes the requirement that the gravitational action involves just two derivatives, and requires only a sensible higher derivative expansion, the constraint on the large N value of the central charges is weakened to \cite{Buchel:2008vz}
\begin{equation}
c,a \gg1 \, , \hspace{1cm} \frac{|c-a|}{c}\ll 1 \,.
\label{buchelagain}
\end{equation}
Going through the list of theories considered here, we observe that this condition is satisfied precisely by the $\nu=0$ theories. It seems that having a Wigner eigenvalue density is necessary to have a gravitational description with a sensible higher derivative expansion.

After this holographic interlude, we come back to the task of solving the saddle-point equation (\ref{thesaddleeq}).

\subsection{Infinite coupling limit}
In the strict limit $\frac{1}{\lambda}=0$, $\mu\rightarrow\infty$ and equation (\ref{thesaddleeq}) reduces to 
\begin{equation}
\int_{-\infty}^\infty dy \rho(y)\left(\frac{1}{x-y}-\nu K(x-y)\right)=-\nu K(x) \, .
\end{equation}
This equation can be solved analytically for $\nu \neq 0$. Taking its Fourier transform we arrive at
$$
\hat \rho_\infty(p)=\frac{1}{1+\frac{2}{\nu}\sinh ^2\frac{p}{2}} \, ,
$$
which implies
$$
\rho_\infty(x)=\frac{1}{\sqrt{\frac{2}{\nu}-1}}\frac{\sinh\left((\pi-\theta)x\right)}{\sinh \pi x} \, ,
$$
with
\be
\theta=\cos ^{-1} (1-\nu) \, .
\label{deftheta}
\ee
This result is just a slight generalization of the $\nu=1$ case, already obtained in \cite{Passerini:2011fe}.

\subsection{Strong coupling}
At finite coupling, we are not aware of a technique that allows to solve the saddle-point equation, (\ref{thesaddleeq}). For finite but strong 't Hooft coupling, $\lambda \gg 1$, there are a couple of works in the literature using different approximations to solve this equation. We will follow \cite{Passerini:2011fe} and also briefly comment on the approximation used in \cite{Bourgine:2011ie}.

The first approach to solve approximately the saddle point equation (\ref{thesaddleeq}) will closely follow  \cite{Passerini:2011fe}, and it is based in the Wiener-Hopf method. Our computations will only differ in the treatment of the zero-momentum mode.

Given the integral equation (\ref{thesaddleeq}), one might be tempted to solve it via a Fourier transform, after extending the definition of $\rho(x)$ to be zero outside its support, $[-\mu,\mu]$. This idea cannot be implemented to (\ref{thesaddleeq}) as it stands, since the Fourier transforms of $K(x)$ and $x$ are divergent. To arrive at an equation amenable to be Fourier transformed, we follow \cite{Passerini:2011fe} and 
make use of the integral operator
$$
\mathcal{P}^{-1}_{x\rightarrow z}\left[f\left(x\right)\right]=-\frac{1}{\pi^2}\fint_{-\mu}^{\mu}\frac{dx}{z-x}\sqrt{\frac{\mu^{2}-z^{2}}{\mu^{2}-x^{2}}}f\left(x\right)
$$
which inverts the principal part integral operator in the following regard:
$$
\mathcal{P}^{-1}_{x\rightarrow z}\left[\fint_{-\mu}^{\mu}dy\frac{\rho\left(y\right)}{x-y}\right]=
\rho\left(z\right);\;z\in\left[-\mu,\mu\right]
$$
Its action onto (\ref{thesaddleeq}) leads to
\be
\rho\left(z\right)-\frac{8\pi}{\lambda}\sqrt{\mu^{2}-z^{2}}-\nu\int_{-\mu}^{\mu}dy\rho\left(y\right)\left(f\left(y,z\right)-f\left(0,z\right)\right)=0;\:z\in\left[-\mu,\mu\right]
\label{modifsaddle}
\ee
$$
f\left(y,z\right)\equiv\mathcal{P}^{-1}_{x\rightarrow z}\left[K\left(x-y\right)\right]
$$
The kernel does not only depend on the difference $z-y$ anymore, so the use of Fourier transformation would lead now to more involved integral expressions. 
We observe nonetheless that by virtue of the symmetry $y\leftrightarrow-y$ the result (\ref{modifsaddle}) will remain valid if we use
$$
\hat{f}\left(y,z\right)\equiv\mathcal{P}^{-1}_{x\rightarrow z}\left[\fint_{-\infty}^{\infty}\frac{\omega\coth\left(\pi\omega\right)}{x-y-\omega}\right]=\left(z-y\right)\coth\left(\pi\left(z-y\right)\right)+\delta \hat{f}\left(y,z\right)
$$
in place of $f\left(y,z\right)$. The advantage in this replacement is that the Fourier transform of the term $\delta \hat{f}\left(y,z\right)$ can be argued to be small, and therefore subdominant in the saddle point equation. This endows us with the possibility of solving the equation iteratively, using at each step the distribution obtained in the previous iteration to improve the estimate on the term that contains $\delta \hat{f}$. For our purposes the first step of the algorithm suffices, where this subleading term is fully neglected.

Once we have reformulated the original equation in this fashion, we are finally ready to apply the Wiener-Hopf method. The first step is to extend the definition of the eigenvalue density $\rho(y)$, outside the interval $[-\mu,\mu]$, by defining $\rho(y)=0$ outside this interval. 
This is compatible with analytic methods for $\rho(y)$ as long as it is understood that $\rho(y)$ admits a  branch cut outside the domain of integration and we are taking the ill-defined values on it as
$$
\rho\left(\left|x\right|>\mu\right)=\frac{1}{2}\lim_{\epsilon\rightarrow 0} \left(\rho\left(x+i \epsilon\right)+\rho\left(x-i\epsilon)\right)\right)
$$
Provided that we take the Fourier transform of the eigenvalue density with this prescription, we obtain 
$$
\int_{-\infty}^{\infty} e^{-i p z}\left(
\hat \rho(p)\left(1+\frac{\nu}{2\sinh^2 \frac{p}{2}}\right)
-F\left(p\right)
\right)=0;\;z\in\left[-\mu,\mu\right]
$$
$$
\int_{-\infty}^{\infty} e^{-i p z}\hat \rho\left(p\right)=0;\;z\not\in\left[-\mu,\mu\right]
$$
$$
F(p)\equiv 8\pi^2 \mu\frac{J_1\left(\mu p\right)}{\lambda p}+\frac{\nu}{2\sinh^2 \frac{p}{2}}+\ldots
$$
where the dots make reference to the terms coming from $\delta\hat{f}$ that we are neglecting. The general solution for the Fourier transform of the eigenvalue density should consequently be of the form
\be
\hat \rho(p)\left(1+\frac{\nu}{2\sinh^2 \frac{p}{2}}\right)= F\left(p\right)-\chi_{-} \left(p\right)-\chi_{+} \left(p\right) \, ,
\ee
where the functions $\chi_{\pm}$ in the position space are nonvanishing in the real line only on one side of $\left|x\right|>\mu$ each. Their exact expressions can be determined from analiticity constraints in momentum space.

In order to impose those constraints we should pause our calculation for a moment to focus on the analytic structure of
$$
1+\frac{\nu}{2\sinh^2 \frac{p}{2}} \, .
$$
This function does have double poles at $p=2 \pi n i$ and simple zeroes at $p=2 \pi n i \pm \theta$ with $\theta$ defined in eq. (\ref{deftheta}). The following splitting will turn out to be very convenient:
$$
1+\frac{\nu}{2\sinh^2 \frac{p}{2}}\equiv\frac{1}{G_+ \left(p\right)G_- \left(p\right)}
$$
$$
G_{+} \left(p\right)\equiv \frac
{p^2 \Gamma \left(1+\frac{\theta-i p}{2 \pi}\right) 
\Gamma \left(1-\frac{\theta+i p}{2\pi}\right)}
{\left(p+i\theta\right)\Gamma^{2} 
\left(1-\frac{ip}{2\pi}\right)}
$$
because the constructions
$$
C_+ =\frac{\hat \rho \left(p\right) e^{-i p \mu}}{G_- \left(p\right)};\;
C_- =\frac{p^2 \hat \rho \left(p\right) e^{i p \mu}}{G_+ \left(p\right)};
$$
are either totally annihilated or left invariant by the action of $\int_{-\infty}^{\infty}\left(2\pi i\right)^{-1}\left(p-p_0\pm i\epsilon\right)^{-1}$ operators. We can straightforwardly read expressions for $\chi_\pm$ from these projections. We obtain
\be
\hat{\rho}\left(p\right)=
\frac{2\sinh^{2}\frac{p}{2}}{2\sinh^{2}\frac{p}{2}+\nu}
\left(F\left(p\right)-\frac{e^{i p \mu}}{G_{+}\left(p\right)}
\sum_{\mbox{$\alpha\in$poles $G_{+}$}}\frac{e^{-i\alpha\mu}F\left(\alpha\right)R_{\alpha}}{p-\alpha}\right)+\mathcal{O}\left(e^{-i p \mu}\right)
\label{therhoinp}
\ee
$$
R_{\alpha}\equiv \hbox{Res}\left(G_{+},\alpha\right)
$$

The expression we have obtained for $\hat{\rho}$ is only useful to obtain $\rho\left(x\right)$ at $x\gg-\mu$, but this covers our needs in this case because of the $x\leftrightarrow-x$ symmetry. The normalization condition can be applied as
$$
1=2\int_{0}^{\infty}dx\rho\left(x\right)=\lim_{\epsilon \rightarrow 0}\frac{1}{i\pi}\int_{-\infty}^{\infty}dp\frac{\hat{\rho}\left(p\right)}{p-i\epsilon}
$$
\be
0=\sum_{\begin{array}{c}\mbox{$\alpha\in$ poles $G_{+}$} \\ \mbox{$\beta\in$ poles $G_{-}$}\end{array}} \frac{e^{-i\left(\alpha-\beta\right)\mu}F\left(\alpha\right)R_{\alpha}\tilde{R}_{\beta}}{\beta-\alpha}
\ee
$$
R_{\alpha}\equiv \hbox{Res}\left(G_{+},\alpha\right);\;\tilde{R}_{\beta}\equiv \hbox{Res}\left(G_{-},\beta\right)
$$
Observe that $F\left(\alpha\right)$ has an exponential contribution that makes all $\alpha$ poles equally important, but the sum in $\beta$ will be dominated by the pole at $\beta=i\theta$. Keeping only this dominant contribution and using asymptotic expressions for the Bessel functions in $F\left(p\right)$ we obtain an equation for the dependence $\mu\left(\lambda\right)$, which at large $\lambda$ can be summarized as
\be
\theta\mu=\ln\lambda-\frac{1}{2}\ln\mu+\mathcal{O}\left(1\right)
\label{normalization}
\ee
The expression for the eigenvalue density (in momentum space), eq. (\ref{therhoinp}) together with the normalization (\ref{normalization}) are the main result of this section. 

Before we put these results to work, let's briefly comment on a different approximation to solve the saddle-point eq. (\ref{thesaddleeq}). In \cite{Bourgine:2011ie}, Bourgine solved (\ref{thesaddleeq}) by truncating the expansion of $K(x)$ and keeping only the first terms in a large $x$ expansion,
\be
K(x) \rightarrow K_{sc}(x) =2x\ln |x|+2\gamma x+\frac{1}{6x} 
\ee
This truncation simplifies the computation enormously, compared with the method we just described. As explained in \cite{Bourgine:2011ie}, when computing the vev of the Wilson loop, it works remarkably well in capturing the exponent, but not so well with the prefactor. For the sake of comparison, the expressions work out to be the same, with the replacement
\be
\theta_B= \sqrt{\frac{2\nu}{1-\frac{\nu}{6}}}
\ee
Remarkably, this expression differs from $\theta=\cos^{-1}(1-\nu)$ in less than 1,8$\%$ in the range $0\leq \nu\leq 1$. Presumably, keeping further terms in the large $x$ expansion of $K(x)$ would improve the agreement of these two methods. Nevertheless, we will stick to the results obtained by the first method, since they capture exactly the exponent in the power law dependence of $\vev{W}$.

\section{Results}
In this section we put to use the eigenvalue densities found in the previous section, by computing various quantities that characterize the heavy probe. We first compute the vacuum expectation value of the Wilson loop itself; we then compute the normalized two-point function of the Wilson loop and the Lagrangian density, and similarly the normalized two-point function of the Wilson loop and the stress-energy tensor. From this last result we deduce the Bremsstrahlung function and the entanglement entropy associated to the probe.

\subsection{Circular Wilson loop}
We start by computing the vev of a 1/2-BPS circular Wilson loop. In his seminal paper \cite{Pestun:2007rz}, Pestun showed that due to localization, the path integral reduces to a matrix model. In the saddle-point approximation, the integral boils down to a rather simple expression in terms of the eigenvalue density,
\be
\vev{W}=\int_\Gamma e^{2\pi x} \rho(x) dx \, .
\ee
When $\nu=0$, the eigenvalue density follows the semi-circle law, and the vev of the Wilson loop displays exponential growth \cite{Berenstein:1998ij, Erickson:2000af}
\be
\rho_G(x;\lambda)= \frac{4}{\lambda}\sqrt{\lambda-(2\pi x)^2}\ \ \ \ \ ; \ \ \ \ \ \langle W(C) \rangle=\frac{2}{\sqrt{\lambda}}I_1(\sqrt{\lambda})\sim\frac{e^{\sqrt{\lambda}}}{(\lambda)^{3/4}}\nonumber
\ee
On the other hand, for any given $\nu \neq0$, we obtain that the vev of the circular Wilson loop displays a power law dependence on $\lambda$
\be
\theta \mu(\lambda) +\frac{1}{2}\ln \mu (\lambda)\sim \log(\lambda) \ \ \ \ \ ;\ \ \ \ \ \langle W(C)\rangle \sim \frac{\sqrt{\mu(\lambda)}}{\lambda}e^{2\pi\mu(\lambda)}\sim \left(\frac {\lambda}{\sqrt{\log(\lambda)}} \right) ^{\frac{2\pi}{\theta}-1}\nonumber
\ee
For $\nu=1$ we have $\theta =\frac{\pi}{2}$ and we recover the known result, $\vev{W}_{\nu=1}\sim \lambda^3$ \cite{Passerini:2011fe}. For the other value of $\nu$ realized by Lagrangian theories, $\nu=1/2$ we have $\theta=\frac{\pi}{3}$ and we obtain
\be
\vev{W}_{\nu=1/2}\sim \lambda^5 \,.
\ee
It is amusing that for the two values of $\nu$ realized by large N Lagrangian ${\cal N}=2$ CFTs, $\nu=1/2$ and $\nu=1$, the exponent in the power law dependence of $\vev{W}$ happens to be given by integers. We don't know if there is any deeper reason behind this observation.

\subsection{Two-point function of the Lagrangian density and the Wilson loop}
We now want to compute the normalized two-point function of the 1/2 BPS Wilson loop and the Lagrangian density. We will first derive a general expression for such two-point function, valid for any Lagrangian CFT, and then evaluate it for the theories at hand.

Consider any CFT that can be written in terms of a Lagrangian density. The Lagrangian density is a scalar operator with scaling dimension $\Delta = 4$. Conformal invariance fixes the normalized two-point function with a straight Wilson line to be
\be
\frac{\vev{{\cal L}(x)W}}{\vev{W}}=\frac{f_W(g^i)}{|\vec x|^4}
\label{defoff}
\ee
where the coefficient $f_W(g^i)$ is a function of the possible marginal couplings of the theory.

For any  Euclidean CFT, a conformal transformation maps the straight Wilson line to a circular one. It is well-known that there is a conformal anomaly associated with this mapping, and the vacuum expectation values of these two operators do not coincide \cite{Erickson:2000af, Drukker:2000rr}. Nevertheless, the contribution of this anomaly is localized on the Wilson line, so it is reasonable to  expect  that it cancels in a normaliized two-point function like the one above, and the same coefficient $f$ also appears in a similarly normalized two-point function with the circular Wilson loop. This expectation is borne out by explicit computations \cite{Giombi:2006de, Gomis:2008qa, Fiol:2012sg,  Buchbinder:2012vr}.

We are going to write an expression for $f_W$ in terms of the vev of  the circular Wilson loop. To do so, we are going to assume that by field redefinitions we can write the action in such a way that the gauge coupling appears only as an overall factor. The vev of the Wilson loop is
\be
\vev{W}=\frac{\int {\cal D}\phi W e^{-\frac{1}{g^2} \int d^4 x {\cal L}}}{\int {\cal D}\phi e^{-\frac{1}{g^2} \int d^4 x {\cal L}}} \, ,
\ee
and we have
\be
g^2\partial_{g^2}\ln \vev{W}=-\frac{1}{g^2}\int d^4x \frac{\vev{{\cal L}(x)W}}{\vev{W}} \, .
\ee
This gives us a relation in terms of the integrated two-point function. To proceed we have to do the integral in the numerator, which is divergent. A convenient regularization was used in \cite{Lewkowycz:2013laa}. It consists of mapping the space to  $S^1\times H_3$, 
\be
ds^2=d\tau^2+d\rho^2+\hbox{sinh} ^2 \rho \left(d\theta^2+\sin^2\theta d\phi^2\right)
\ee
and introduce a short distance cut-off $\rho_c$ for the coordinate $\rho$. The divergence appears then as a pole $1/\rho_c$, which is discarded. Following this procedure we arrive at
\be
f_W=\frac{1}{8 \pi^2} g^2\partial_{g^2} \ln \vev{W}
\label{thelagcoef}
\ee

This expression is valid for any Lagrangian  4d CFT, supersymmetric or not. As a check, for ${\cal N}=4$ SYM, this relation coincides, up to a number, with the expression found for the Bremsstrahlung coefficient in \cite{Correa:2012at}, 
\be
4 f_W^{{\cal N}=4} = B^{{\cal N}=4} \, ,
\ee
and these coefficients must indeed be related in this way, since on the one hand, in ${\cal N}=4$, the lagrangian density and the stress-energy tensor are in the same supermultiplet, and on the other hand, for ${\cal N}=4$ theories, the Bremsstrahlung function is related to the two-point function of the stress-energy tensor and the Wilson loop \cite{Fiol:2012sg, Lewkowycz:2013laa}. 

Having derived a general formula for this coefficient, we can now use the results just derived for $\vev{W}$ to obtain this coefficient for ${\cal N}=2$ SCFTs, in the large N, large $\lambda$ regime. For theories with $\nu=0$, we reproduce the known result \cite{Callan:1999ki},
\be
f_W=\frac{\sqrt{ \lambda}}{16 \pi^2} \, .
\ee

For theories with $\nu \neq 0$, we find that at large $\lambda$ and large N, the leading term in $f(\lambda)$ is independent of $\lambda$
\be
f_W=\frac{1}{8 \pi^2}\left(\frac{2 \pi}{\theta}-1\right) \, .
\ee
In our derivation, this result follows immediately from the fact that the Wilson loop grows only as a power law for large $\lambda$. Nevertheless, we find it quite remarkable. From its definition (\ref{defoff}) we can interpret this coefficient as giving the strength of the fields sourced by a static probe; our computation implies that for superconformal theories with matter in the fundamental representation, this strength reaches a limiting value in the large N, large $\lambda$ limit.

\subsection{Two-point function of the stress-energy tensor and the Wilson loop}
We move now to the computation of a similarly normalized two-point function, that of the stress-energy tensor and the Wilson loop. Again, for a straight Wilson line, conformal invariance fixes this two-point function up to a coefficient \cite{Kapustin:2005py},
\be
\frac{\vev{T_{00}W}}{\vev{W}}=\frac{h_W(g^i)}{|\vec x|^4}\, .
\ee
It was recently conjectured \cite{Fiol:2015spa} that for ${\cal N}=2$ SCFTs, this coefficient can be related to the vev of a circular Wilson loop in a squashed four-sphere $\mathbb{S}^4_b$ \cite{Alday:2009fs,Hama:2012bg}, since varying the squashing parameter will insert the stress-energy tensor,
\be
h_W=\frac{1}{12 \pi^2} \partial_b \ln \vev{W_b}|_{b=1}\, .
\label{thebguess}
\ee

Furthermore, it was argued in \cite{Fiol:2015spa} that this computation can be carried out by just inserting $W_b$ in the matrix model for $\mathbb{S}^4$,
\be
\vev{W_b}=\int \; dx \; e^{2\pi b x} \rho(x) \, .
\ee
which is a computation we can readily perform using the results derived in the previous section. For $\nu=0$  we have $\vev{W_b}=e^{b\sqrt{ \lambda}}$ so applying eq. (\ref{thebguess}) we arrive at
\be
h_W=\frac{\sqrt{ \lambda}}{12 \pi^2}\, ,
\label{hwinfour}
\ee
a result that can be alternatively obtained by a supergravity computation \cite{Friess:2006fk}. 
 
For $\nu\neq 0$ theories, it is more convenient to compute $\vev{W_b}$ directly in momentum space using
$$
\vev{W_{b}}=\hat{\rho}\left(-2\pi b i\right) \, .
$$
Keeping the relevant term in the asymptotic $\lambda\gg 1$ limit and plugging (\ref{normalization}) into the result we conclude that
\be
\ln\left\langle W\right\rangle _{b}\sim\ln\lambda\left(\frac{2\pi b}{\theta}-1\right)+\mathcal{O}\left(\left(1-b\right)^2,\frac{\ln \mu}{2} \right)\, 
\ee
so we arrive at
\be
h_W=\frac{1}{6\pi \theta}\ln \lambda \,.
\ee
Again, the result for any $\nu\neq 0$ differs parametrically from the known result of $\nu=0$ theories, which displays the ubiquitous $\sqrt{\lambda}$ dependence, as in  eq. (\ref{hwinfour}). Notice also that for generic ${\cal N}=2$ theories, the $\lambda$ dependence of the two coefficients just considered is different. This should not come as a surprise, since for ${\cal N}=2$ theories (unlike what happens in ${\cal N}=4$ SYM) the Lagrangian density and the stress-energy tensor don't belong to the same supermultiplet.

As a bonus, the computation of $h_W$ immediately gives us two other interesting quantities. The first one is the Bremsstrahlung function of the corresponding probes. For any 4d CFT, the Bremsstrahlung coefficient can be defined \cite{Correa:2012at} as the coefficient that appears in the formula for the energy loss of an accelerated probe,
\be
E=2\pi B\int \; dt\; a^2 \, .
\ee
It also captures the momentum fluctuations of the accelerated probe \cite{Fiol:2013iaa}. Intuitively, it seems natural that the two-point function of the stress-energy tensor would capture the energy loss of the probe. However, the details are subtle and there is no simple universal relation for B and $h_W$ \cite{Lewkowycz:2013laa}, valid for all four-dimensional CFTs. Nevertheless, for probes of ${\cal N}=2$ SCFTs it is conjectured  \cite{Lewkowycz:2013laa, Fiol:2015spa} that
\be
B=3h_W \, .
\ee
and granting that this conjectured relation is true, we conclude that
\be
B=\frac{1}{2\pi \theta}\ln \lambda \, .
\ee
One lesson of this result is the following. It has been argued in \cite{Pomoni:2013poa,Mitev:2014yba} that a certain class of observables of planar ${\cal N}=2$ superconformal gauge theories can be obtained from the corresponding result of planar ${\cal N}=4$ SYM, by means of replacing the ${\cal N}=4$ coupling by a single function, universal for a given ${\cal N}$=2 SCFT. Comparing the results we have obtained for $\vev{W}$ and $B$ for ${\cal N}$=4 and ${\cal N}$=2 theories, we conclude that this substitution rule does not apply to the computation of $B$, for theories with a single gauge group.

Finally, we can use our result for $h_W$ to compute the additional entanglement entropy of a spherical region when we add a external probe to the vacuum of the theory. According to \cite{Lewkowycz:2013laa} it is given by
\be
S=\ln\vev{W}-8\pi^2h_W \, ,
\ee
so for the probes we are considering we have
\be
S=\left(\frac{2\pi}{3\theta}-1\right)\ln \lambda \, .
\ee

\acknowledgments
{We would like to thank Efrat Gerchkovitz, Zohar Komargodski, Vladimir Mitev and Elli Pomoni for discussions. The research of BF and GT is partially supported by the Spanish MINECO under projects FPA2013-46570-C2-2-P and MDM-2014-0369 of ICCUB (Unidad de Excelencia ``Mar\'ia de Maeztu"), and by AGAUR, grant 2014-SGR-1474. GT is further supported by an FI scholarship by the Generalitat de Catalunya.

\bibliographystyle{newutphys}
\bibliography{ProbeLocal}

\end{document}